\documentclass[preprint,preprintnumbers,amsmath,amssymb,color]{revtex4}

\usepackage{graphicx}
\usepackage{bm}

\def\o1{{\mathrm{o}(1)}}
\begin{document}

\preprint{CAS-KITPC/ITP-158}

\title{~\\ \vspace{2cm}
Casimir Energy, Holographic Dark Energy and Electromagnetic
Metamaterial Mimicking de Sitter Space \vspace{1cm}}

\author{Miao Li}\email{mli@itp.ac.cn}
\author{Rong-Xin Miao}\email{mrx11@mail.ustc.edu.cn}
\author{Yi Pang}\email{yipang@itp.ac.cn}
\affiliation{Kavli Institute for Theoretical Physics, Key Laboratory
of Frontiers in Theoretical Physics, Institute of Theoretical
Physics, Chinese Academy of Sciences, Beijing 100190, People's
Republic of China, Interdisciplinary Center for Theoretical Study,
University of Science and Technology of China, Hefei, Anhui 230026,
China\vspace{2cm}
}%

\begin{abstract}
We compute the Casimir energy of the photon field in a static de
Sitter space and find it to be proportional to the size of the
horizon, the same form of the holographic dark energy. We suggest to
make metamaterials to mimic static de Sitter space in laboratory and
measure the predicted Casimir energy.

\end{abstract}


\maketitle

\section{Introduction}

The dark energy problem is one of the central unsolved problems in
fundamental physics and cosmology. Many phenomenological models have
been proposed since the discovery of the accelerated expanding
universe \cite{cc}, including the so-called holographic dark energy
model \cite{Li}.

The value of the observed dark energy is comparable to the critical
energy density which in the Planck unit is proportional to the
Hubble constant squared. Motivated by this fact, the holographic
dark energy model presumes that the energy density is inversely
proportional to the square of the size of some cosmological horizon,
and the choice of the horizon, for the model to work, is proposed in
\cite{Li} to be the event horizon. In a pure de Sitter space in
which dark energy is a constant, one can choose a static coordinate
system, thus the total energy of dark energy can be defined. The
total energy in the Planck unit is proportional to the size of the
horizon, unlike the usual result for the Casimir energy in the
finite cavity in a flat spacetime, the latter is inversely
proportional to the size of the cavity. Nevertheless, there has been
no attempt to computing the Casimir energy in a static de Sitter
space seriously, to the best of our knowledge.

In this paper, we take the assumption that the UV divergent
zero-point energy is somehow regularized to vanish due to some
unknown mechanism, therefore the observed dark energy is solely due
to the residual zero-point energy which is known to be the Casimir
energy. However, we are faced with the challenge that why the
Casimir energy is not inversely proportional to the size of the
cavity.

Motivated by the recent development in electromagnetic
metamaterials, we will first try to convert the de Sitter space to a
cavity of optical metamaterial, this is doable at least for the
photon field. We shall find that the cavity is drastically different
from a usual one in that the permittivity and permeability
parameters both have divergent components tangent to the boundary,
indeed these components are divergent near the boundary which is
derived from the horizon of the de Sitter space. This fact
encourages us to guess that the Casimir energy as a function of the
size of the cavity behaves differently from that in a normal cavity.
We carry out the calculation for the photon field and find that
indeed after regularization there is still a divergent piece
proportional to the size of the cavity, with the divergence cut-off
by a UV scale. This UV scale in the metamaterials can be some
intrinsic microscopic scale, and becomes the Planck scale when
transformed back to the de Sitter space. Thus, we find a piece of
theoretical evidence for the holographic dark energy model. We
propose to do some laboratory experiment to verify our result about
the Casimir energy. This kind of experiment is interesting in two
aspects. First, it will mimic cosmology, second, it will detect a
unusual Casimir energy inversely proportional to some microscopic
cut-off.

There are two different designs of metamaterials, the first is based on
the static de Sitter space with the usual coordinate system, the second
is based on the static de Sitter where the radial coordinate is replaced
by the proper distance. The Casimir energy in the second metamaterial
assumes the precisely same form as in the de Sitter space, which is proportional
to the size of the material thus is enhanced to be easily detectable.

In the next section, we present the Maxwell equations in a curved
spacetime. We discuss how to transform the de Sitter space into a
cavity of metamaterial in sect.3 and compute the Casimir energy in
sect.4, and discuss its form in a metamaterial filled cavity in
sect.5. We conclude in sect.6.

\section{Maxwell equation in static curved space time}
We start with Maxwell equations in a curved spacetime
\begin{equation}{\label{MX1}}
 \partial_{[\mu}F_{\nu\lambda]}=0 ,\mbox{ }\mbox{ }\mbox{
 }\partial_{\mu}H^{\mu\nu}=0,
\end{equation}
where the first equation is the Bianchi identity satisfied by field
strength $F_{\mu\nu}$, and $H^{uv}$ is defined by
\begin{equation}
H^{\mu\nu}=\sqrt{-g}F^{\mu\nu}=\sqrt{-g}g^{\mu\alpha}g^{\nu\beta}F_{\alpha\beta}.
\end{equation}
The electric field $E_{i}$ and magnetic field $H_i$ are related to
$F_{\mu\nu}$ as
\begin{equation}
 E_{i}=F_{0i}, \mbox{ }\mbox{
 }H_{i}=-\frac{\epsilon_{ijk}}{2}H^{jk},
\end{equation}
In above equations, the Levi-Civita symbol $\epsilon^{ijk}$ is $+1$
for all even permutations of $\epsilon^{123}$. It should be noted
that $E_{i}$ and $H_{i}$ are spatial vectors.

Especially, when the space time is static, we can express Eq.
(\ref{MX1}) in a spatial covariant form with respect to optical
metric $\gamma^{ij}$ defined by
\begin{equation}
\gamma^{ij}=-g_{00}g^{ij},
\end{equation}
where $\gamma^{ij}$ is the usually called optic metric.

At this stage, Maxwell equations Eq.(\ref{MX1}) can
be rewritten as
\begin{equation}\label{MX6}
 \nabla_{i}E^{i}=0,\mbox{ }\mbox{ }\mbox{ }\nabla_{i}H^{i}=0,
\end{equation}
\begin{equation}{\label{MX4}}
 \partial_{t}E^{i}-\frac{\epsilon^{ijk}}{\sqrt{\gamma}}\partial_{j}H_{k}=0,\mbox{ }\mbox{
 }\partial_{t}H^{i}+\frac{\epsilon^{ijk}}{\sqrt{\gamma}}\partial_{j}E_{k}=0,
\end{equation}
where $\nabla_{i}$ denotes the covariant derivative with respect to
$\gamma^{ij}$, and all the indices are raised and lowered by
$\gamma^{ij}$.

Eq.(\ref{MX4}) can be rewritten in the following compact form
\begin{equation}
 i\partial_{t}\left (\begin{array}{l}E\\H
\end{array}\right)=L\left (\begin{array}{l}E\\H
\end{array}\right),
\end{equation}
where the operator $L$ is given by
\begin{equation}\label{maxwell}
 L=i\left
(\begin{array} {ccc}0& \gamma_{ij}\frac{\epsilon^{jkl}}{\sqrt{\gamma}}\partial_{k}\\
 -\gamma_{ij}\frac{\epsilon^{jkl}}{\sqrt{\gamma}}\partial_{k}& 0\\
\end{array}\right).
\end{equation}

If we define
\begin{equation}\label{}
    \phi=\left (\begin{array}{l}E\\H
\end{array}\right),
\end{equation}
the operator $L$ is Hermitian in terms of the following inner
product
\begin{equation}
 <\phi_{1},\phi_{2}>=\int d^{3}x \sqrt{\gamma}\phi^{+}_{1}\phi_{2}=\int
 d^{3}x(E_{1i}\varepsilon^{ij}E_{2j}+H_{1i}\mu^{ij}H_{2j}),
\end{equation}
where
\begin{equation}\label{opticalmetric}
    \varepsilon^{ij}=\mu^{ij}=\sqrt{\gamma}\gamma^{ij},~~~~~~~\gamma=\mbox{det}(\gamma_{ij}).
\end{equation}
We will see that $\epsilon$ and $\mu$ have a physical interpretation in terms of metamaterials in
the next section.

To make $L$ Hermitian under the inner product (10) (which is the
energy functional as it should be for $\phi_{1}=\phi_{2}$), we need
to specify boundary conditions if there is a boundary. We find
 the following boundary conditions
\begin{equation}\label{BC}
 \sqrt{\gamma}(E_{1}\times H_{2}-H_{1}\times E_{2})^i n_i|_{\partial M} =0
\end{equation}
where  $n_i$ is the normal vector of boundary and $(E\times
H)^{i}=\frac{1}{\sqrt{\gamma}}\epsilon^{ijk}E_{j}H_{k}$. Then the
hermiticity of $L$ is guaranteed
\begin{equation}
 <\phi_{1},L\phi_{2}>=<L\phi_{1},\phi_{2}>.
\end{equation}

 By acting $\partial_t$ and $L$ again on the two sides of Eq.(7) respectively, we obtain the equations of electric and magnetic field
\begin{eqnarray}
 \nonumber-\partial^{2}_{t}
 E_{i}=(\nabla^{j}\nabla_{i}-\nabla^2\delta^j_i)E_{j},\\
   -\partial^{2}_{t}
 H_{i}=(\nabla^{j}\nabla_{i}-\nabla^2\delta^j_i)H_{j},
\end{eqnarray}
we read off the operator corresponding to $L^2$ from the above equations
\begin{equation}\label{maxwell}
D\equiv L^2=\left
(\begin{array} {ccc}\nabla^{j}\nabla_{i}-\nabla^2\delta^j_i& 0\\
 0& \nabla^{j}\nabla_{i}-\nabla^2\delta^j_i
\end{array}\right)\\.
\end{equation}
An eigenvalue of $D$ is a nonnegative real number due to the Hermiticity
of $L$.

With Gauss law, we get two sets of equations taking the
same form
\begin{equation}{\label{MX5}}
-\partial^{2}_{t}
 V_{i}=D^j_iV_j=(\nabla^{j}\nabla_{i}-\nabla^2\delta^j_i)V_{j},~~~~~~\nabla^iV_i=0,
\end{equation}
where $V_i$ can be $E_i$ or $H_i$.

\section{Mimicking de Sitter with metamaterials }
Recently, the analogy between static curved space and inhomogeneous
medium has been studied in \cite{Pendry1, Pendry2, Pendry3, Leo}. It
is proved in \cite{NK} that in geometric optic limit, light
propagates along the same path in electromagnetic material and in
static curved spacetime as long as the optic metric $\gamma_{ij}$ is
related to permittivity and permeability by Eq.
(\ref{opticalmetric}).

 One can see the analogy more clearly as follows. To solve Maxwell equation in inhomogeneous
medium conveniently, the often used quantities are electric
displacement field $D^i$ and magnetic induction field $B^i$ defined
as
\begin{equation}
 D^{i}=\varepsilon^{ij}E_{j},\mbox{ }\mbox{ }B^{i}=\mu^{ij}H_{j},
\end{equation}
where $\varepsilon^{ij}$ and $\mu^{ij}$ are permittivity and
permeability specifying the electromagnetic properties of the
medium. In terms of  the newly defined quantities $D^i$ and $H^i$,
Maxwell equations in inhomogeneous medium and Cartesian coordinate
system are
\begin{equation}{\label{MX2}}
 \partial_{i}D^{i}=0,\mbox{ }\mbox{ }\mbox{ }\partial_{i}B^{i}=0,
\end{equation}
\begin{equation}{\label{MX3}}
 \partial_{t}D^{i}-\epsilon^{ijk}\partial_{j}H_{k}=0,\mbox{ }\mbox{
 }\partial_{t}B^{i}+\epsilon^{ijk}\partial_{j}E_{k}=0.
\end{equation}
Since under spatial coordinate transformation $\varepsilon^{ij}$ and
$\mu^{ij}$ are tensor density with weight -1, we can define two
tensors using $\varepsilon^{ij}$ and $\mu^{ij}$ by
\begin{equation}
\tilde{\gamma}^{ij}=\frac{\varepsilon^{ij}}{\mbox{det}(\varepsilon^{ij})},~~~~~~\tilde{\tilde{\gamma}}^{ij}=\frac{\mu^{ij}}{\mbox{det}(\mu^{ij})}.
\end{equation}
For electromaganetic material with equal
$\varepsilon^{ij}$ and $\mu^{ij}$,
$\tilde{\gamma}^{ij}=\tilde{\tilde{\gamma}}^{ij}$. In this case,
rewriting  Eqs.(\ref{MX2}) and (\ref{MX3}) in terms of
$\tilde{\gamma}^{ij}$, $E_i$ and $H_i$, we find that Maxwell equations in
the metameterial
takes the same form as Eq.(\ref{MX6}) and Eq.(\ref{MX4}) if
$\tilde{\gamma}^{ij}$ is identified with the optic metric
$\gamma^{ij}$.

By this analogy, one can design electromagnetic metamaterial
corresponding a given metric. Along this line, theoretically, a new
material called electromagnetic cloak was conceived
\cite{Pendry1,Pendry2,Pendry3,Leo}. All parallel
bundles of incident rays are bent around some region covered by
electromagnetic cloak and recombined in precisely the same direction
as they entered the medium; Experimentally, a new class of
electromagnetic materials are invented \cite{exp} which can be
designed to have properties mimicking novel gravitational effects in Nature.
Thus it is now conceivable that a material can be constructed whose
permittivity and permeability values may be designed to vary
arbitrarily.

With the static de Sitter space
\begin{equation}
 ds^{2}=-(1-\frac{r^{2}}{L^{2}})d^{2}t +
 (1-\frac{r^{2}}{L^{2}})^{-1}d^{2}r + r^{2}d^{2}\Omega
\end{equation}
The permittivity and permeability are
\begin{equation}\label{}
\varepsilon^{rr}=\mu^{rr}=r^2
\sin\theta,~~~~\varepsilon^{\theta\theta}=\mu^{\theta\theta}=\frac{\sin\theta}{1-\frac{r^{2}}{L^{2}}},~~~~
\varepsilon^{\varphi\varphi}=\mu^{\varphi\varphi}=\frac{1}{(1-\frac{r^{2}}{L^{2}})\sin\theta},
\end{equation}
where $(r, \theta, \varphi)$ stands for spherical coordinate. In
terms of Cartesian coordinate, the permittivity and permeability
are
\begin{equation}\label{p1}
\varepsilon^{ij}=\mu^{ij}=\frac{1}{1-r^2/L^2}(\delta^{ij}-\frac{x^ix^j}{L^2}).
\end{equation}

 The metamaterial can always be redesigned by making a coordinate
 mapping \cite{Pendry1, Pendry3}. By making the following mapping
 \begin{equation}\label{}
 r\rightarrow\tilde{r}=L \arcsin(r/L).
 \end{equation}
Then the permittivity and permeability becomes
\begin{equation}\label{}
\varepsilon^{\tilde{r}\tilde{r}}=\mu^{\tilde{r}\tilde{r}}=L^{2}\frac{\sin^2(
\tilde{r}
/L)}{\cos(\tilde{r}/L)}\sin\theta,~~~~\varepsilon^{\theta\theta}=\mu^{\theta\theta}=\frac{\sin\theta}{\cos(\tilde{r}/L)},~~~~
\varepsilon^{\varphi\varphi}=\mu^{\varphi\varphi}=\frac{1}{\cos(\tilde{r}/L)\sin\theta}.
\end{equation}
where $(\tilde{r}, \theta, \varphi)$ denotes the spherical
coordinate. In terms of the Cartesian coordinate
\begin{equation}\label{p2}
\varepsilon^{ij}=\mu^{ij}=\frac{1}{\cos(\tilde{r}/L)}(\delta^{ij}-(\frac{L^{2}}{\tilde{r}^{2}}\sin^{2}(\tilde{r}/L)-1)\mbox{
}\frac{x^ix^j}{ \tilde{r}^2}).
\end{equation}

These two kinds of metamaterial mimicking the same de Sitter space,
since they are related by a spatial coordinate transformation.
However, the different permittivity and permeability will lead to
distinct physical phenomenon in laboratory due to their different
physical composition.

The event horizon at $r=L$ or $\tilde{r}=\pi L/2$ now becomes the
boundary of a cavity of metamaterial. As we shall see shortly, to
metamaterial with $\epsilon$ and $\mu$ as in Eq.(\ref{p1}), the
leading term in the Casimir energy is a constant inversely
proportional to certain microscopic length scale; while to
metamaterial with $\epsilon$ and $\mu$ as in Eq.(\ref{p2}), the
leading term in the Casimir energy in this cavity is proportional to
$L$, we encourage experimenters to design such cavity to measure the
Casimir energy, the result will have important implication for dark
energy.

\section{Casimir energy of Electromagnetic field in static de Sitter space}

In Maxwell theory, the electric sector and magnetic
sector contribute the same amount to the Casimir energy, as the equations
governing them are identical.  In the following, we
will focus on the electric sector.

Consider the solution of electric field taking the form
\begin{equation}\label{solution}
    E_i(t,x)=e^{i\omega t}E_i(x),
\end{equation}
where the absolute value of $\omega$ should be interpretated as the
energy with respect to time $t$. Upon quantization, the eigenvalue
$\omega$ contributes to the zero-point energy. The Casimir
energy of electric field is the infinite sum
\begin{equation}\label{}
    E_{\rm Casimir}=\frac{1}{2}\sum_{\omega}|\omega|.
\end{equation}

Substitutting Eq.(\ref{solution}) into Eq.(\ref{MX5}) we obtain the
eigen equation of operator $D$ with eigenvalue $\omega^2$
\begin{equation}\label{}
    D_{i}^{j}E_{j}=\omega^2E_i.
\end{equation}
Then in the framework of zeta function regularization, the problem
of computing the Casimir energy is converted into the problem of
computing the heat kernel of $D$, the Casimir Energy can be
extracted from the heat kernel as follows \cite{Blau}:
\begin{equation}\label{CE}
E_{\rm Casimir}\equiv \lim_{\epsilon\rightarrow0}
\frac{1}{2}\{E_{\rm reg}(+\epsilon)+E_{\rm reg}(-\epsilon)\},
\end{equation}
where
\begin{equation}
E_{\rm reg}(\epsilon)=\frac{\mu}{2}\mbox{
}\zeta(-\frac{1}{2}+\epsilon),
\end{equation}
with the zeta function associated with $D$ defined by
\begin{equation}\label{zeta}
 \zeta(s)=\frac{\mu^{2s}}{\Gamma(s)}\int_{0}^{\infty}dt \mbox{ }t^{s-1}
\mbox{tr}^{'}(e^{-tD}).
\end{equation}
In above expressions, a scale ``$\mu$'' is introduced to keep the
zeta function dimensionless; while the prime in $\mbox{tr}^{'}$
indicates that zero eigenvalues are not included. In coordinate
representation, $\langle x|e^{-tD}|x^{'}\rangle$ is called the heat
kernel denoted by $K(t,x,x^{'})$ satisfying the heat equation
related to $D$ as
\begin{equation}\label{HE}
    \partial_tK(t,x,x^{'})+DK(t,x,x^{'})=0.
\end{equation}
To solve this equation, we need to specify appropriate boundary conditions.

  At this moment, with above preparation, we can
   calculate the Casimir energy of electromagnetic field in static de Sitter space time. This space time is
   described by the following metric
\begin{equation}
 ds^{2}=-(1-\frac{r^{2}}{L^{2}})d^{2}t +
 (1-\frac{r^{2}}{L^{2}})^{-1}d^{2}r + r^{2}d^{2}\Omega
\end{equation}
where $L$ is the de Sitter radius. However, the effective metric
relevant to our calculation is the optic metric $\gamma^{ij}$
appearing in the Maxwell equation with the following form
\begin{equation}\label{om}
 ds^{2}_3=\gamma_{ij}dx^idx^j=
 (1-\frac{r^{2}}{L^{2}})^{-1}((1-\frac{r^{2}}{L^{2}})^{-1}d^{2}r +
 r^{2}d^{2}\Omega),
\end{equation}
where the lower index ``3'' is used to emphasize that this metric is
three dimensional. An interesting observation is that $\gamma_{ij}$
actually describes an anti de Sitter space. One can see this clearly
by performing the following transformation on $r$ coordinates:
\begin{equation}\label{}
    r=\frac{\tilde{r}}{\sqrt{1+\tilde{r}^2/L^2}}.
\end{equation}
In terms of $\tilde{r}$,
\begin{equation}
 ds^{2}_3=
 (1+\frac{\tilde{r}^{2}}{L^{2}})^{-1}d^{2}\tilde{r} +
 \tilde{r}^{2}d^{2}\Omega.
\end{equation}
In the case of static de Sitter space time, the boundary condition
Eq.(\ref{BC}) for retaining the Hermiticity of operator $L$ in Eq.(\ref{maxwell}) becomes
\begin{equation}\label{BC1}
    (E_{1\theta}H_{2\phi}-H_{1\theta}E_{2\phi})|_{r=L-\delta}=0,
\end{equation}
where the boundary is chosen at $r=L-\delta$ and $\delta$ is a cut
off to avoid the divergence of energy and its meaning will become
clear later (see later for an explicit calculation of energy). Thus
to keep Hermiticity, we choose
\begin{equation}\label{BC2}
 E_{\theta}|_{r=L-\delta}=E_{\varphi}|_{r=L-\delta}=0
\end{equation}
as the boundary condition. It is similar to the case of
electromagnetic field in a spherical conductor shell. Therefore in
the following, we will concentrate on the problem of finding the
heat kernel corresponding to $D$ defined with respect to
$\gamma_{ij}$ in Eq.(\ref{om}) and restricted to the space spanned
by its eigenfunctions satisfying the Gauss law and boundary
conditions.
\begin{equation}\label{BC3}
    \nabla^iE_i=0,~~~~~~~E_{\theta}|_{r=L-\delta}=E_{\varphi}|_{r=L-\delta}=0.
\end{equation}
Usually, in a curved manifold the heat equation Eq.(\ref{HE}) is
hard to solve completely. However, the trace of the heat kernel has
the following asymptotic expansion for small $t$, especially in
three dimensions as
\begin{equation}\label{}
    \mbox{tr}^{'}(e^{-tD})=(\frac{1}{4\pi t})^{3/2}\{\mbox{
}\sum_{0}^{N}(\int_M a_{n}t^{n}+\int_{\partial M}
b_{n}t^{n})+o(t^{N})\mbox{ }\},
\end{equation}
where $a_n$ are bulk terms independent of boundary condition,
composed by polynomial of Riemann tensor, Ricci tensor, Ricci
scalar, and their covariant derivatives; $b_n$ are  boundary terms
comprised of the curvature, second fundamental form and their
derivatives on the boundary. The sum runs over half-integers, but
$a_n$ vanishes for half-odd-integers. Above asymptotic expansion
implies that $\zeta(s)$ has a pole structure \cite{Blau}:
\begin{equation}\label{}
\zeta(s)=\frac{\mu^{2s}}{\Gamma(s)(4\pi)^\frac{3}{2}}\{\sum_{n=0}^{\infty}\frac{a_n+b_n}{s+n-\frac{3}{2}}+f(s)\},
\end{equation}
where $f(s)$ is an entire analytic function of $s$, but in general,
one has little information about it \cite{Blau}. In the previous
version of this work, we calculate the Casimir energy with above
formulas. It is found that the leading term comes from a surface
integral on the boundary horizon. Based on this expression, one may
think that the Casimir energy is localized near the boundary
horizon, but this may be a misunderstanding. For instance, in the
frame work of heat kernel expansion, the Casimir energy between two
infinite conductor plane in flat space will turn up as a boundary
term depending on the permittivity and permeability of the conductor
appearing to be localized on the boundary. However, L. Ford's result
\cite{Ford}
 shows that in this
case the Casimir energy distributes homogeneously between the two
conductor planes.

To avoid this misunderstanding, We will seak for a compact
expression for the heat kernel. It is recalled that the optical
metric describes an Euclidean $\rm{AdS}_3$ space and the heat kernel
defined with respect to it can be solved by utilizing Harmonic
analysis \cite{Camporesi:1990wm}. To vector field the corresponding
heat kernel satisfying the transverse condition and vanishing at the
$\rm{AdS}_3$ boundary was given by \cite{David:2009xg} (In our
convention, the $\rm{AdS}_3$ boundary corresponds to $r=L$ or
$\tilde{r}=\infty$). From Eq. (\ref{BC3}), it is obvious that the
result of \cite{David:2009xg} amounts to the $\delta\rightarrow 0$
limit of the heat kernel in our case. In other words, we can use the
result of \cite{David:2009xg} as a good approximation to the heat
kernel sought by us up to the order $\delta/L$. Based on that result
we write down the contracted coincident limit of the heat kernel as

\begin{equation}\label{}
    \gamma^{ij}K_{ij}(x,x,t)=\frac{2}{(4\pi
    t)^{\frac{3}{2}}}(1+2\frac{t}{L^2})e^{-t/L^2}+o(\delta/L).
\end{equation}
With above result, $\rm{tr}^{'}(e^{-tD})$ can be evaluate by

\begin{eqnarray}
\nonumber \rm{tr}^{'}(e^{-tD}) &=& \int\sqrt{\gamma}\gamma^{ij}K_{ij}(x,x,t)d^3x \\
       &=& \pi L^3 (\frac{L}{\delta}-\ln(\frac{2L}{\delta}))\times\frac{2}{(4\pi
    t)^{\frac{3}{2}}}(1+2\frac{t}{L^2})e^{-t/L^2}+o(1),
\end{eqnarray}
where $o(1)$ denotes terms $\ll \frac{L}{\delta}$. Substituting this
result into Eqs. (\ref{zeta}) and (\ref{CE}), we obtain
\begin{equation}\label{}
E_{\rm
Casimir}=\frac{3}{16\pi}(\ln\mu^2-\gamma-\psi(-\frac{1}{2}))(\frac{1}{\delta}-\frac{1}{L}\ln(\frac{2L}{\delta}))+o(1/L),
\end{equation}
where $\gamma$ is Euler constant and
$\psi(-\frac{1}{2})=\Gamma^{'}(-\frac{1}{2})/\Gamma(-\frac{1}{2})$.
In above result, $\delta$ is a coordinate cut off so the physical
meaning is not clear. To express it in terms of the physical cut
off, we recall that in static de Sitter space the physical distance
between $r=L-\delta$ and $r=L$ is given by
\begin{equation}
\Delta l_{\rm  phy }=L \arcsin(r/L)|^{r=L}_{r=L-\delta}.
\end{equation}
By requiring $\Delta l_{\rm phy }$ to be the Planck length $l_{\rm
p}$, we derive
\begin{equation}\label{}
    \delta=L(1-\cos(\frac{l_{\rm
p}}{L}))\approx\frac{l_{\rm p}^2}{2L^2}.
\end{equation}
In terms of the physical cut off, the Casimir energy of
electromagnetic field is

\begin{equation}\label{}
E_{\rm
Casimir}=\frac{3}{8\pi}(\ln\mu^2-\gamma-\psi(-\frac{1}{2}))(\frac{L}{l_{\rm
p}^2}-\frac{1}{L}\ln(\frac{2L}{l_{\rm p}}))+o(1/L),
\end{equation}
where the dominant term proportional to $L/l^{2}_{p}$ takes the same
form as in the holographic dark energy model \cite{Li}. Compared
with previous result given in \cite{Candelas:1978gf,Birrell}, it is
found that the leading term in the vacuum energy density of static
de Sitter space is proportional to $(L^2-r^2)^{-2}$ which can also
generate a leading term proportional to $L/l^{2}_{\rm{p}}$ after
integrated over the bulk of static de Sitter space. Finally we note
that the running renormalization $\ln \mu^2$ can be absorbed into
$l_{\rm{p}}^2$ or be determined experimentally \cite{Birrell}.

\section{Casimir energy of Electromagnetic field in Electromagnetic metamaterial mimicking de Sitter}

We have designed two kinds of metamaterial mimicking de Sitter in
sect.3. For an electromagnetic metamaterial with $\epsilon$ and
$\mu$ taking the value of Eq.(\ref{p1}) in laboratory, coordinate
$r$ has been the physical distance between origin and $r$, since
metric measuring the physical distance is flat. In this case, the
physical IR cut off is chosen to be $L-d$, and the Casimir energy is
given by
\begin{equation}\label{CE3}
    E_{\rm
Casimir}=\frac{3}{16\pi}(\ln\mu^2-\gamma-\psi(-\frac{1}{2}))(\frac{1}{d}-\frac{1}{L}\ln(\frac{2L}{d}))+o(1/L),
\end{equation}
where $d$ is the counterpart of the Planck length in metamaterial,
its value depends on the detailed chemical components and structure
of the material which should be determined by experiment. We note
that when $L\gg d$,
\begin{equation}
   E_{\rm
Casimir}\approx
\frac{3}{16\pi}(\ln\mu^2-\gamma-\psi(-\frac{1}{2}))\frac{1}{d}.
\end{equation}
 $L$ is usuaully much larger than $d$, the
leading term is much larger than the usual term proportional to
$1/L$ (for example, if $d$ is  nanometer, and $L$ is 1cm, then the
leading term is about $10^6$ times of the second term, the usual
term which is almost undetectable).

Alternatively, to metamaterial with $\epsilon$ and $\mu$ taking the
form as in Eq.(\ref{p2}), the Casimir energy is

\begin{equation}\label{}
    E_{\rm
Casimir}=\frac{3}{8\pi}(\ln\mu^2-\gamma-\psi(-\frac{1}{2}))L^{-1}(\frac{\sin(\tilde{r}/L)}{\cos^2(\tilde{r}/L)}-\frac{1}{2}\ln(\frac{1+\sin(\tilde{r}/L)}{1-\sin(\tilde{r}/L)}))|_{
\tilde{r}=\pi L/2 -d \rightarrow \pi
 L/2}+o(1/L),
\end{equation}
where $\tilde{r}$ stands for physical distance between origin and
$\tilde{r}$ in this coordinate system and the size of the cavity is
$\pi L/2$. So distinguished from the first case, $\pi L/2 -d$ is the
physical IR cut off. $d$ has the same meaning as in previous case.
Usually $L\gg d$
\begin{equation}\label{}
   E_{\rm
Casimir}\approx\frac{3}{8\pi}(\ln\mu^2-\gamma-\psi(-\frac{1}{2}))\frac{L}{d^2}.
\end{equation}

It is remarkable that in this metamaterial, the Casimir energy has
the same form as in its gravity analogy, except that some
microscopic scale $d$ takes the place of Planck scale.
\section{Conclusion}

It is a surprising result that the Casimir energy in a de Sitter
space is not inversely proportional to the size of the horizon as
for a usual finite cavity, rather it is proportional to the size of
the horizon. However, if one is to expect that the Casimir energy is
dark energy or at least a part of dark energy, this is a desired
result, the normal answer in a cavity is just too small to be
relevant in cosmology.

Although we only computed the Casimir energy due to the photon
field, it is expected that other fields will have the same form of
Casimir energy. Since we suggest to construct a cavity of
metamaterial to do experiment in laboratory, it is enough to know
the form of the photon field Casimir energy, since this is what to
be measured in laboratory. However, because the corresponding
Casimir force is combined with the elastic force, it seems difficult
to measure the predicted Casimir energy immediately.

Aside from the issue of dark energy, to use metamaterials to mimic cosmology by itself is an exciting
future direction, we look forward to developments in the future.

\section*{Acknowledgments}

We would like to thank Prof. Mo Lin Ge for informing us of the exciting developments
in the field of electromagnetic cloaking and metamaterials. This work is mostly
motivated by looking for the possibility of studying cosmology in laboratory with metamaterials.
This work was supported by the NSFC grant
No.10535060/A050207, a NSFC group grant No.10821504 and Ministry of
Science and Technology 973 program under grant No.2007CB815401.

\end{document}